# Thermoelectric properties of $Pr_3Rh_4Sn_{13}$-type $Yb_3Co_4Ge_{13}$ and $Yb_3Co_4Sn_{13}$ compounds

A.V. Morozkin[a], V.Yu.Irkhin[b], V.N.Nikiforov*[c]

[a]Department of Chemistry, Moscow State University, Leninskie Gory, House 1, Building 3, Moscow, GSP-2, 119992, Russia

[b]Institute of Metal Physics, Ural Division of the Russian Academy of Sciences, Ekaterinburg, Russia

[c]Physics Department, Moscow State University, Leninskie Gory, Moscow, GSP-2, 119899, Russia

E-mail: nvn@lt.phys.msu.ru

## Abstract

Crystallographic data and thermoelectric properties (from 240 K up to 380 K) of $Yb_3Co_4Ge_{13}$, $Yb_3Co_4Sn_{13}$ compounds and $Yb_2CeCo_4Ge_{13}$ and $Yb_{2.3}La_{0.7}Co_4Ge_{13}$ solid solutions are reported. The Seebeck coefficients, electric resistance and thermal conductivity increase monotonically with increasing temperature from 240 to 380 K for all the compounds. The Seebeck coefficient is $S = 14 \div 27$ μV//K for $Yb_3Co_4Sn_{13}$, and $S = -21 \div -12$ μV/K for $Yb_3Co_4Ge_{13}$. The substitution of Yb for cerium or lanthanum in $Yb_3Co_4Ge_{13}$ shifts the Seeebeck coefficient to positive values. The $Yb_3Co_4Sn_{13}$ has a maximal ZT parameter from available $Pr_3Rh_4Sn_{13}$-type compounds. The ZT parameter of $Yb_3Co_4Sn_{13}$ compound increases from 0.006 up to 0.017 with increasing temperature from 240 K to 380 K.

## Introduction

The interest in skutterudite-related systems is connected with search of new thermoelectric materials [1]. Unfilled skutterudites of the type $M(P,Sb,As)_3$ (M is a transition metal) contain voids into which low-coordination ions, in particular rare earth elements, can be inserted. This increases phonon scattering and decreases lattice thermal conductivity K without increasing electrical resistitivity ρ [2]. Thus the figure of merit $Z = S^2/(\rho K)$ (S is the Seebeck coefficient) can become rather large.

The electronic properties of the $Ir_4LaGe_3Sb_9$, $Ir_4NdGe_3Sb_9$, and $Ir_4SmGe_3Sb_9$ systems were investigated in Ref.[2]. Another class of skutterudite-related systems are $R_3T_4X_{13}$ compounds (where R is a rare earth element, T = Ge or Sn) which attract attention due to the interesting electron properties, interplay of superconductivity and magnetic order (see [3]). In particular, the Yb-based systems demonstrate the intermediate valence nature of the Yb ions and slightly enhanced γ value of the electronic specific heat. Physical properties and superconductivity of skutterudite-related $Yb_3Co_{4.3}Sn_{12.7}$ and $Yb_3Co_4Ge_{13}$ were considered in Ref. 3.

Our paper reports the crystal structure and thermoelectric properties of the $Yb_3Co_4Sn_{13}$ and $Yb_3Co_4Ge_{13}$ compounds and $Yb_{3-x}R_xCo_4Ge_{13}$ (R = Ce, La) solid solutions.

## Experimental details

In the present investigation the compounds were prepared in an electric arc furnace under argon atmosphere using the non-consumable tungsten electrode and a water-cooled copper tray. Germanium, tin (purity all components 99.99%), ytterbium, lanthanum, cerium and cobalt (purity components 99.9%) were used as the starting components. For melting process, zirconium was used as a getter. Subsequently, the compounds were annealed at 820-1070 K for 170 h in an argon atmosphere and quenched in ice-cold water.

The quality of the samples before physical measurements was determined using X-ray phase analysis and microprobe X-ray analysis. X-ray data were obtained on a diffractometer DRON-3.0 (Cu $K_\alpha$ radiation, $2\Theta$ = 20-70 deg, step 0.05 deg, for 5 s per step). The diffractograms obtained were identified by means of calculated patterns using the Rietan-program [4] in the isotropic approximation. A "Camebax" microanalyzer was employed to perform microprobe X-ray spectrum and SEM images of the samples.

Samples in the shape of rectangular parallelepiped with typical dimensions 1.0 x 1.0 x 4 $mm^3$ were cut from the ingots using a wire saw.

The electrical resistance along the long axis of the sample was measured with a standard four-terminal geometry using a dc current of 0.8 mA (we did not detect a sample overheating effect at this current magnitude). To attach copper wires (0.14 mm in diameter) to the sample the electrical contacts were made using the low-temperature conductive epoxy (Lake Shore Cryotronics, Inc.). Contact resistances were typically few Ohm. To exclude the parasitic thermoelectric power in the potential circuit, the current direction was changed, and two voltage reading were added. Typical voltage across the sample was few microvolts, while parasitic thermopower could be as large as 20 microvolts. In order to reduce a random experimental error, averaging over the 36 successive readings was applied. An error in determination of absolute value of the electrical resistivity was approximately 20%.

The Seebeck coefficient (the thermoelectric voltage under zero electric current) has been measured with disconnection of the sample from the current source. The heat flux in the sample was generated with electric heater. The heater made from a chip resistor of 820 Ohm was attached to the sample with the conductive epoxy, but was not applied as electric contact to the sample. The thermovoltage E was measured over the same potential contacts used in the resistance measurements. The temperature drop dT between these contacts was measured with differential manganin-constantan thermocouple which was glued to the copper leads with GE7031 varnish. This thermocouple has been made from thin wire of manganin (7 micrometers) and constantan (30 micrometers). The Seebeck coefficient S was calculated as the ratio E/dT. The uncertainty of S was estimated to be less than 10%.

The thermal conductivity was measured by a longitudinal steady state method. The temperature drop over the sample was generated by the electric heater. The sample was clamped at another end to a copper base. Heat losses from the sample were minimized by evacuating a sample chamber. The temperature of the copper base was measured with a standard type E thermocouple (chromel-constantan). The temperature drop along the sample was measured using the differential manganin-constantan thermocouple. Both manganin and constantan have low values of thermal conductivity and therefore this thermocouple does not disturb significantly the heat flow pattern in the sample. The magnitude of the temperature drop was approximately 0.5–2 K. The random error in the magnitude of thermal conductivity was about 3%. The systematic error was less than 20%. The latter is due mostly to error in determination of the sample dimensions, irregularity of its form, finite size of the contact between thermocouple junctions and the sample, and uncertainty in the magnitude of heat current through the sample.

## Results and Discussion

The $R_3T_4X_{13}$ compounds (R = Rare Earths, T = Ru, Rh, Os, X = Ge, Sn) adopt the $Pr_3Rh_4Sn_{13}$-type structure [5-9] (some related problems and controversies are discussed in Ref [3]). In this structure (space group Pm3n No. 223) the rare-earth atoms occupy the special position 6(d) (1/4, 1/2, 0), transition metal atoms are located in the position 8(c) (1/4, 1/4, 1/4), germanium (tin) atoms are in 2(a) (0, 0, 0) and 24(f) (0, $Y_{X2}$, $Z_{X2}$) positions [6]. The structure has a cage-type nature: two 12-coordination cages surround X1 atoms and R atoms, and the third cage is centered on the empty position (see Fig.1 and Figure 1 of Ref.[9]).

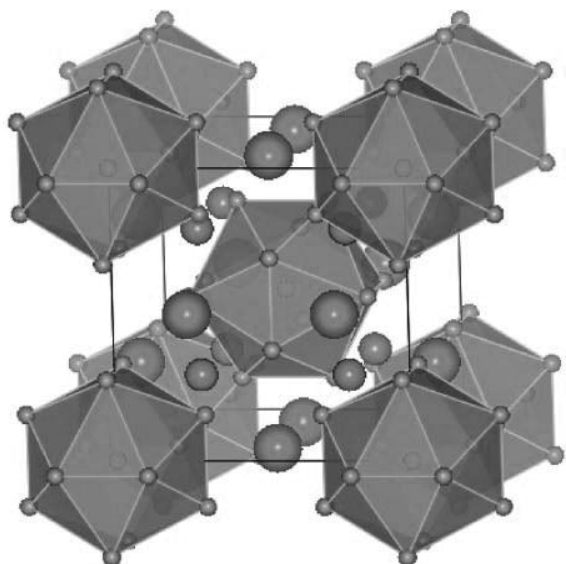

Fig.1. Crystal structure of $R_3T_4X_{13}$ compounds according to Ref. [8]. The $X$1 atoms are surrounded by icosahedral clusters from twelve $X$2 atoms (small circles). The $T$ atoms are shown by middle-size circles, $R$ atoms by large circles.

Our analysis of the powder X-ray diffractograms showed that the $Yb_3Co_4Ge_{13}$, $Yb_3Co_4Sn_{13}$, $CeYb_2Co_4Ge_{13}$ and $Yb_{2.3}La_{0.7}Co_4Ge_{13}$ compounds crystallize in the cubic $Pr_3Rh_4Sn_{13}$-type structure (space group Pm3n No. 223). The cell parameters of the compounds, refined at room temperature, atomic position parameters and the reliability factor $R_F$ are given in Table 1.

Table 1.

Cell parameter a (nm) and atomic position parameters of $Pr_3Rh_4Sn_{13}$-type compounds. The R factors are given in percent ($R_F = 100 \bullet (\Sigma_k |(I_k^{obs})^{1/2} - (I_k^{cal})^{1/2}|)/\Sigma_k |(I_k^{obs})^{1/2}|$ %), $I_k^{obs}$ is the integrated intensity evaluated from summation of contribution of the k th peaks to net observed intensity, $I_k^{cal}$ is the integrated intensity calculated from refined structural parameters.

| Compound | a | $Y_{X2}$ | $Z_{X2}$ | $R_F$ |
|---|---|---|---|---|
| $Yb_3Co_4Ge_{13}$ | 0.87295(9) | 0.3155(7) | 0.1578(7) | 3.8 |
| $Yb_2CeCo_4Ge_{13}$ | 0.8747(1) | 0.3168(9) | 0.1556(9) | 5.3 |
| $Yb_{2.3}La_{0.7}Co_4Ge_{13}$ | 0.8744(1) | 0.322(1) | 0.156(1) | 6.7 |
| $Yb_3Co_4Sn_{13}$ | 0.9520(1) | 0.3067(8) | 0.1561(8) | 5.7 |

Inter-atomic distances for $Yb_3Co_4Ge_{13}$ and $Yb_3Co_4Sn_{13}$ compounds are presented in Table 2. Substitutions of germanium for tin and ytterbium for cerium (lanthanum) lead to increasing cell parameters and inter-atomic distances in the $Pr_3Rh_4Sn_{13}$-type compounds.

Table 2

Interatomic distances D with the accuracy of $2 \cdot 10^{-3}$ nm and coordination number N for atoms in $Yb_3Co_4Ge_{13}$ and $Yb_3Co_4Sn_{13}$ compounds.

a). $Yb_3Co_4Ge_{13}$  b). $Yb_3Co_4Sn_{13}$

| Atoms | D(nm) | N | Atoms | D(nm) | N |
|---|---|---|---|---|---|
| Yb – 4Ge2 | 0.3041 | 18 | Yb – 4Sn2 | 0.3318 | 18 |
| - 8Ge2 | 0.3042 |  | - 8Sn2 | 0.3355 |  |
| - 4Co | 0.3086 |  | - 4Co | 0.3366 |  |
| - 2Yb | 0.4365 |  | - 2Yb | 0.4760 |  |
| Co – 6Ge2 | 0.240 | 9 | Co – 6Sn2 | 0.260 | 9 |
| - 3Yb | 0.3086 |  | - 3Yb | 0.3366 |  |
| Ge1–12Ge2 | 0.308 | 12 | Sn1 – 12Sn2 | 0.328 | 12 |
| Ge2 – 2Co | 0.240 | 9 | Sn2 – 2Co | 0.260 | 9 |
| - 1Ge2 | 0.276 |  | - 1Sn2 | 0.297 |  |
| - 2Ge2 | 0.279 |  | - 2Sn2 | 0.316 |  |
| - 1Ge2 | 0.308 | | - 1Sn1 | 0.328 | |
| - 1Yb | 0.3041 | | - 1Yb | 0.3318 | |
| - 2Yb | 0.3042 | | - 2Yb | 0.3355 | |

X-ray phase analysis and microprobe X-ray analysis show that samples contain the admixture phases that may affect the results of the measurements of the thermoelectric properties (Table 3 and Fig. 2).

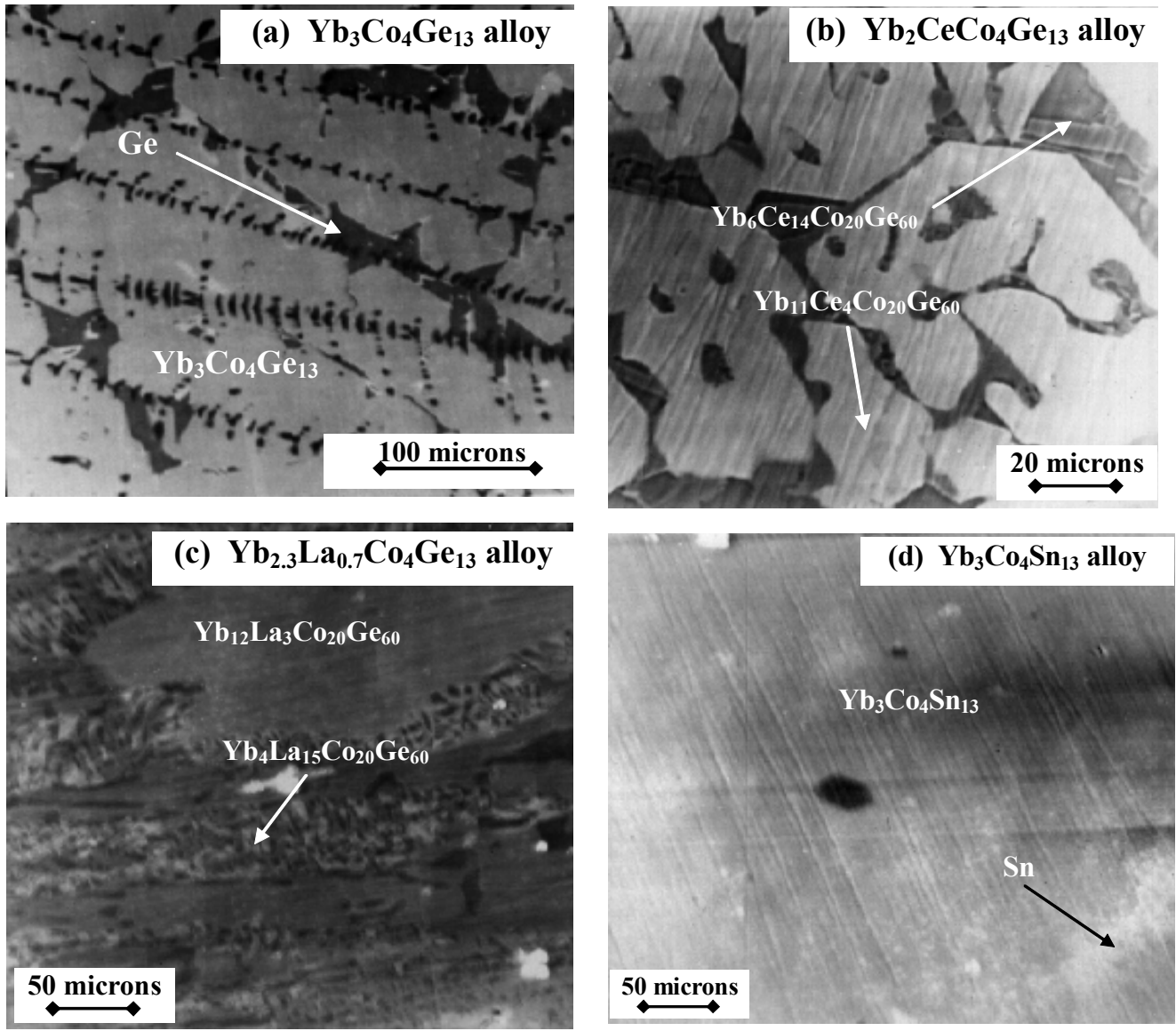

Fig. 2. Scanning electron microscope photos of the $Yb_3Co_4Ge_{13}$ (a), $Yb_2CeCo_4Ge_{13}$ (b), $Yb_{2.3}La_{0.7}Co_4Ge_{13}$ (c) and $Yb_3Co_4Sn_{13}$ (d) alloys.

Table 3.

Mass fraction, composition and crystallographic data for the phases in the $Yb_3Co_4Ge_{13}$, $Yb_2CeCo_4Ge_{13}$, $Yb_{2.3}La_{0.7}Co_4Ge_{13}$ and $Yb_3Co_4Sn_{13}$ alloys. The R factors are given in percent ($R_F = 100 \, (\Sigma_k|(I_k^{obs})^{1/2} - (I_k^{cal})^{1/2}|)/\Sigma_k|(I_k^{obs})^{1/2}|)$ %, $I_k^{obs}$ is the integrated intensity evaluated from summation of contribution of the k th peaks to net observed intensity, $I_k^{cal}$ is the integrated intensity calculated from refined structural parameters.

| Alloys | Phase* | Mass fraction | Type structure | Space group | $a$, nm | $c$, nm | $R_F$, % |
|---|---|---|---|---|---|---|---|
| $Yb_3Co_4Ge_{13}$ | $Yb_{15}Co_{20}Ge_{60}$ | 0.90 | $Pr_3Rh_4Sn_{13}$ | Pm3n | 0.87295(9) | | 3.8 |
| | $Ge_{100}$ | 0.10 | C | Fd3m | 0.5645(2) | | 5.1 |
| $Yb_2CeCo_4Ge_{13}$ | $Yb_{11}Ce_4Co_{20}Ge_{60}$ | 0.87 | $Pr_3Rh_4Sn_{13}$ | Pm3n | 0.8747(1) | | 5.3 |
| | $Yb_6Ce_{14}Co_{20}Ge_{60}$ | 0.13 | $BaNiSn_3$ | I4mm | 0.4295(1) | 0.9814(1) | 7.1 |
| $Yb_{2.3}La_{0.7}Co_4Ge_{13}$ | $Yb_{12}La_3Co_{20}Ge_{60}$ | 0.92 | $Pr_3Rh_4Sn_{13}$ | Pm3n | 0.8744(1) | | 6.7 |
| | $Yb_4La_{15}Co_{20}Ge_{60}$ | 0.08 | $BaNiSn_3$ | I4mm | 0.4317(1) | 0.9860(2) | 7.0 |
| $Yb_3Co_4Sn_{13}$ | $Yb_{15}Co_{20}Sn_{60}$ | 0.83 | $Pr_3Rh_4Sn_{13}$ | Pm3n | 0.9520(1) | | 5.7 |
| | $Sn_{100}$ | 0.17 | Sn | $I4_1/amd$ | 0.5829(1) | 0.3179(1) | 4.7 |

* Data of microprobe X-ray analysis

Meantime, we suppose that the results of the measurements correspond to the $Pr_3Rh_4Sn_{13}$-type compounds, because the impurity phases form the "islands" in the samples (Fig. 2).

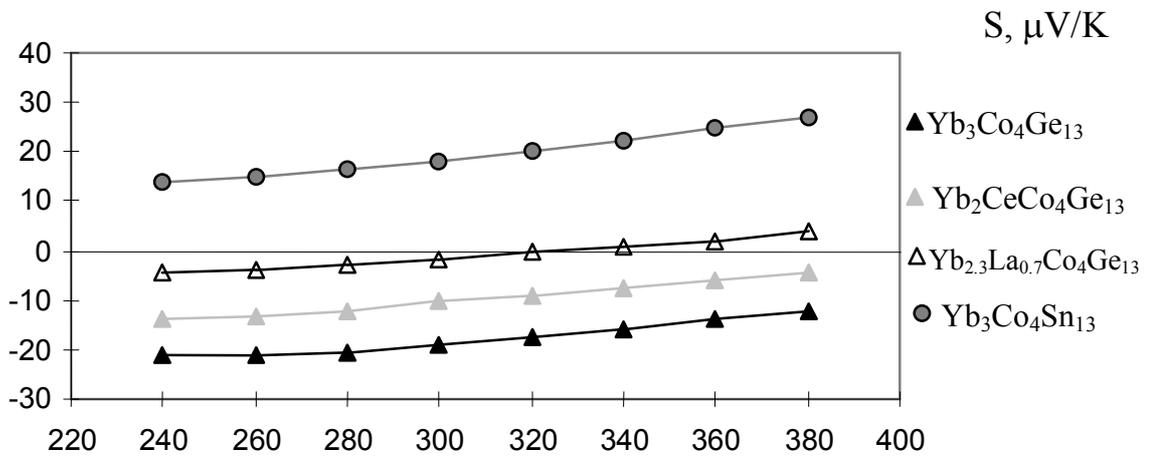

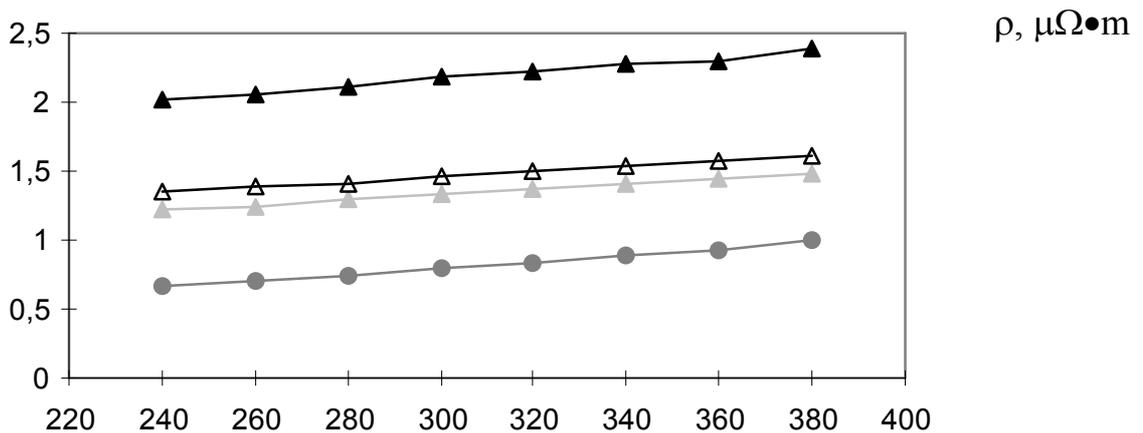

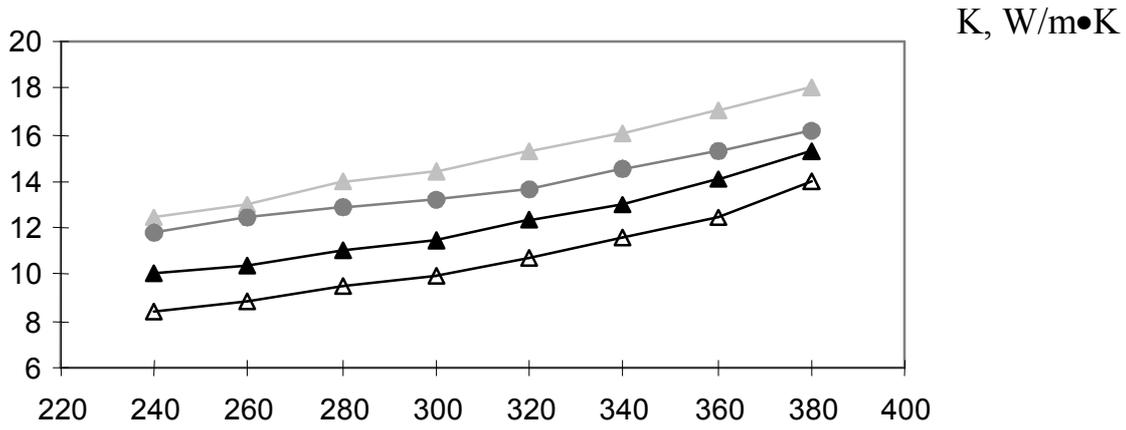

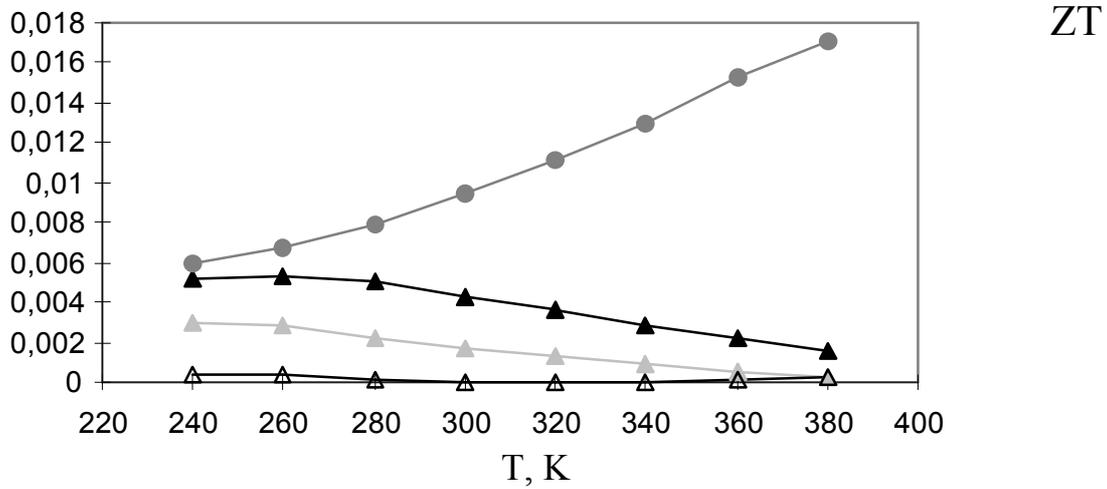

Fig. 3. The Seebeck coefficient (S), electrical resistivity (ρ), thermal conductivity (K) and figure of merit parameter ZT vs. temperature for $Yb_3Co_4Ge_{13}$ (black triangles ), $Yb_2CeCo_4Ge_{13}$ (grey triangles), $Yb_{2.3}La_{0.7}Co_4Ge_{13}$ (white triangles) and $Yb_3Co_4Sn_{13}$(circles) compounds

The Seebeck coefficients, electric resistance and thermal conductivity have monotonic slope with increasing temperature from 240 to 380 K (Fig.3).

All the compounds demonstrate a metallic-type conductivity. Probably, the $Yb_3Co_4Ge_{13}$ has higher electroresistance due to presence of Ge semiconductor in the alloy: the electroresistance of the $Yb_2CeCo_4Ge_{13}$ and $Yb_{2.3}La_{0.7}Co_4Ge_{13}$ alloys is less than of $Yb_3Co_4Ge_{13}$ alloy which demonstrates absence of the Ge admixture phase (Table 3).

It is obvious that the lattice thermal conductivity $K_{lattice}$ of Ge-containing $Pr_3Rh_4Sn_{13}$-type compounds is higher than for Sn-containing compound (thermal coductivity K is the sum of $K_{lattice}$ lattice conductivity $K_{lattice}$ and $K_{electronic}$ electronic conductivity, $K = K_{lattice} + K_{electronic}$) due to that the mass of Sn is higher than mass of Ge. Also, due to this, the Wiedemann-Frantz parameter $WF = \rho K/\gamma_{VF} T$ has higher values for Ge-containing compounds (Fig. 4). Here $\gamma_{VF}$ is the Wiedemann-Frantz constant, $\gamma_{VF} = (\pi^2/3)(k_B/e)^2$, $k_B$ being the Boltzmann constant and e the electronic charge.

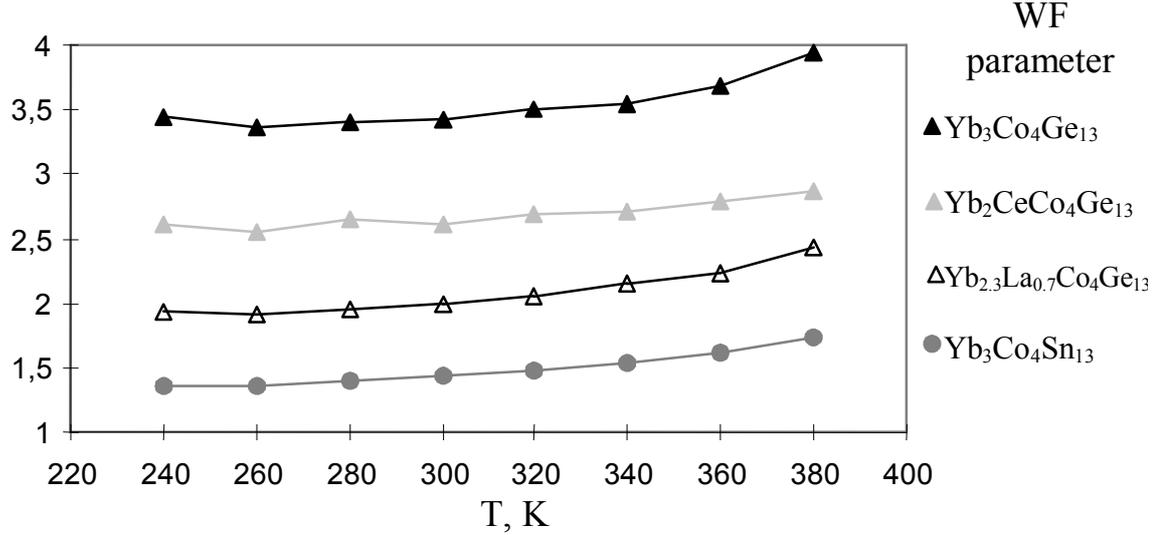

Fig. 4. The Wiedemann-Frantz parameter WF vs. temperature for $Yb_3Co_4Ge_{13}$, $Yb_2CeCo_4Ge_{13}$, $Yb_{2.3}La_{0.7}Co_4Ge_{13}$ and $Yb_3Co_4Sn_{13}$ compounds.

Substitution of Yb for Ce or La, as well as substitution of Ge for Sn, shifts the Seebeck coefficient to positive values observed in the $Yb_3Co_4Ge_{13}$ compound. The Seebeck coefficient of $Yb_{2.3}La_{0.7}Co_4Ge_{13}$ demonstrates a non-monotonous temperature dependence. This sign change may be connected with a non-trivial electronic structure near the Fermi level, characteristic for Kondo-like and intermediate-valence systems (see, e.g., [10]). It should be noted that recent investigations of $R_3T_4X_{13}$ systems have revealed Kondo and heavy-fermion features of Ce-based stannides and germanides [11,12]. The electrical resistivity of $Ce_3Co_4Sn_{13}$ can be treated in terms of a combination of semiconductor-like behavior and Kondo effect [13].

The cage-type structure of $R_3T_4X_{13}$ systems (Fig.1) provides promising thermoelectric properties. Indeed, highly-coordinated atom in a large cage of neighbor atoms can scatter heat-carrying acoustic phonons through off-centre thermal motion suppressing thereby the thermal conductivity K [9]. Besides that, the internal disorder in R and T sites is favorable for low K.

Recently, thermoelectric properties of the compounds $R_3Ru_4Ge_{13}$ (R=Y, Dy, Ho, and Lu) were investigated in [14]. The Seebeck coefficient at room temperature makes up approximately 40 μV/K. However, these compounds display a semiconductor-like behavior of electrical resistivity which is much larger than that of typical metals. Large thermoelectric power S = 38 μV/K at room temperature was also reported for the compound $Y_3Ir_4Ge_{13}$ which is classified as a semimetallic system on the edge of metallic behavior [9].The corresponding value of ZT is 0.02.

Among the systems under consideration in the present work, the parameter $ZT = TS^2/(\rho K)$ is maximal for $Yb_3Co_4Sn_{13}$ compounds and acquires the value of 0.017 at 380 K (Fig. 3). Although its thermal conductivity is not maximal, this compound demonstrates lowest resistivity at high temperatures. Thus a good metallic electrical resistivity and high Seebeck coefficient (rather than small heat conductivity) turn out to be decisive for large *ZT* values in this case.

## Conclusions

We can conclude as follows:

1. The $Pr_3Rh_4Sn_{13}$-type Ge-containing compounds have lattice thermal conductivity higher than Sn-containing compounds due to relatively low Ge atom mass as compared to Sn, which is important for thermoelectric characteristics.

2. The systems investigated demonstrate a non-trivial behavior vs. alloy composition. Substitution of Yb for cerium or La shifts the Seebeck coefficient to positive values. A more systematic treatment is of interest. Probably, some $Yb_{3-x}R_xCo_4Sn_{13}$ compounds will have the Seebeck coefficient and ZT parameter value higher than $Yb_3Co_4Sn_{13}$ compound.

3. Provided that the dependences of electric resistance, thermal conductivity and Seebeck coefficient vs. temperature retain for $Yb_3Co_4Sn_{13}$ compound up to 1000 K, this compound may have ZT value about 0.3 at 1000 K (WF parameter about 3).

Although ZT values obtained are still not too large, the rare-earth compounds are promising materials. There exist some ways to improve their characteristics. Further experimental investigations in this directions and comparison with other ternary rare-earth systems would be useful.

## Acknowledgments

This work was supported by Asahi Kasei Corporation (Japan) in the ISTC project N 2382p and by the Programs of fundamental research of RAS Physical Division "Strongly correlated electrons in solids and structures", project No. 12-T-2-1001 (Ural Branch) and of RAS Presidium "Quantum mesoscopic and disordered structures", project No. 12-P-2-1041,